\pdfoutput=1
\documentclass[aps,twocolumn,amsmath,amssymb,floatfix]{revtex4}
\usepackage{graphicx}
\usepackage{dcolumn}
\usepackage{bm}
\usepackage{amsfonts}
\usepackage{graphicx}
\usepackage{dcolumn}
\usepackage{bm}
\usepackage{amsmath}
\usepackage{amssymb}
\usepackage{cancel}
\usepackage{epstopdf}

\begin{document}
\title{Large negative differential transconductance in multilayer graphene: \\
The role of intersubband scattering}
\author{Seungchan Woo$^{1}$}
\author{E. H. Hwang$^{2}$}
\email{euyheon@skku.edu}
\author{Hongki Min$^{1}$}
\email{hmin@snu.ac.kr}
\affiliation{$^1$ Department of Physics and Astronomy and Center for Theoretical Physics, Seoul National University, Seoul 08826, Korea}
\affiliation{$^2$ SKKU Advanced Institute of Nanotechnology and Department of Physics, Sungkyunkwan University, Suwon, 16419, Korea}

\begin{abstract}
We calculate the transport properties of multilayer graphene, considering the effect of multisubband scattering in a high density regime, where higher subbands are occupied by charge carriers. To calculate the conductivity of multilayer graphene, we use the coupled multiband Boltzmann transport theory while fully incorporating the multiband scattering effects. We show that the allowed scattering channels, screening effects, chiral nature of the electronic structure, and type of impurity scatterings determine the transport behavior of multilayer graphene. We find that the conductivity of multilayer graphene shows a sudden change when the carriers begin to occupy the higher subbands, and therefore a large negative differential transconductance  appears as the carrier density varies. These phenomena arise mostly from the intersubband scattering and the change in the density of states at the band touching density. Based on our results, it is possible to build novel devices utilizing the large negative differential transconductance in multilayer graphene.
\end{abstract}

\maketitle

\section*{INTRODUCTION}
\begin{figure*}[t]
\includegraphics[width=1\linewidth]{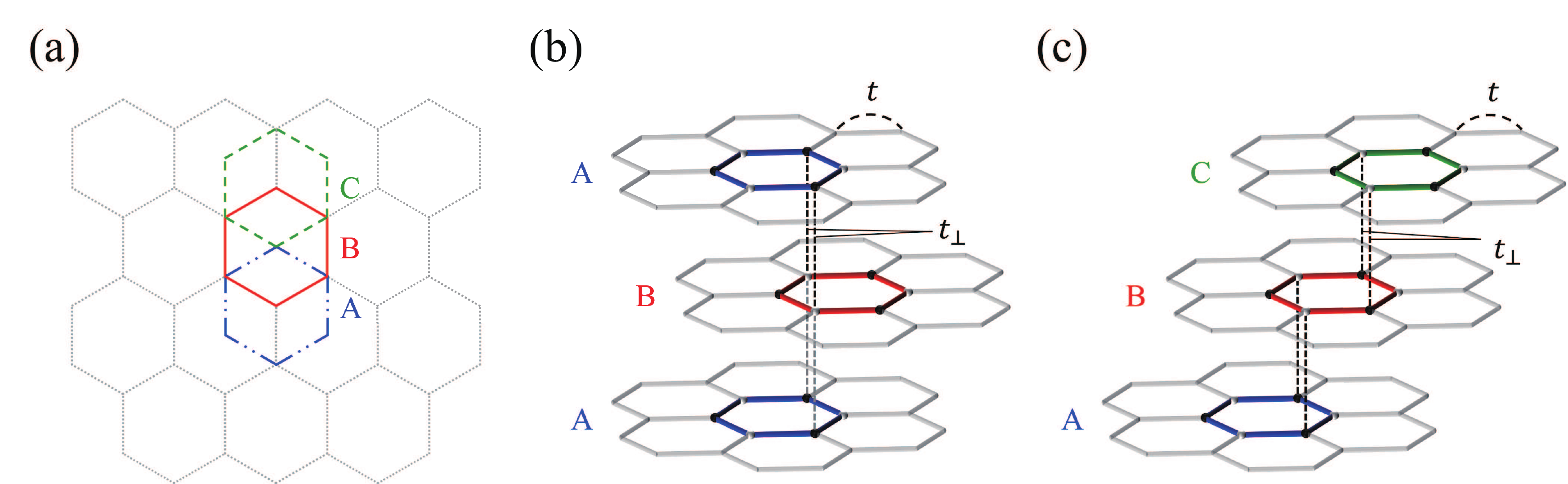}
\caption{(a) Schematic illustration of three inequivalent stacking arrangements in honeycomb lattice, labeled by A, B and C. The neareast neighbor intralayer ($t$) and interlayer ($t_\perp$) 
hopping terms in (b) ABA-stacked and (c) ABC-stacked trilayer graphenes.
}  
\label{fig0}
\end{figure*}

Since its discovery in 2004, graphene has continued to attract considerable attention because of its high carrier mobility and the gate tunability of its charge carrier density.  Recently, there has been an increasing interest in the physics of multilayer graphene and its application to novel electronic and optical devices \cite{HassanRaza2012}. Multilayer graphene is composed of a series of two-dimensional hexagonal lattices of carbon atoms, and its electronic and transport properties strongly depend on the stacking arrangements of each layer (see Fig.~\ref{fig0}). Multilayer graphene shows a distinctive band structure that is different from that of single layer graphene. For example, bilayer graphene is a tunable band gap semiconductor \cite{McCann2006a,McCann2006b,Min2007} and trilayer graphene has a unique electronic structure consisting of massless (linear) and massive (quadratic) subband dispersions \cite{Min2008a,Min2008b}. 

One of the most distinct electronic properties of multilayer graphene compared to single layer graphene is the subband formation due to the interlayer coupling between layers. The energy difference between the subbands is usually on the order of interlayer coupling. Thus, in order to reveal the interesting physics related to the higher subbands, it is necessary to increase the carrier density up to the point that the higher subbands are occupied. However, most study on graphene properties has focused on the low carrier density regime ($n < 10^{13}$ cm$^{-2}$) and near the Dirac point \cite{Sarma2011}. Recent experimental developments make it possible to induce charge carrier densities up to $n \sim 10^{14}$  cm$^{-2}$ through polymer electrolyte gating \cite{Efetov2011} and ionic-liquid gating \cite{Ye2011}, thereby demonstrating the unusual transport properties of multilayer graphene.
Even though a monotonic increase in conductivity with carrier density is usually observed in monolayer graphene \cite{Sarma2011,Hwang2007,Tan2007}, nonlinear conductivity behaviors in multilayer graphene have been experimentally reported at high carrier densities \cite{Efetov2011,Ye2011}. The observed nonmonotonic conductivity behavior is closely related to the carrier occupation of the higher subbands in multilayer graphene. However, the detailed electronic mechanisms for the nonlinear transport behavior have not yet been investigated. 

In this paper, we study the electronic transport properties of multilayer graphene, focusing on the effect of multiband scattering at high carrier densities. From a simple tight-binding Hamiltonian, we calculate the DC conductivity within the \emph{coupled multiband} Boltzmann transport theory and relaxation time approximation, considering both charged Coulomb impurities and short-range scatterers (e.g., lattice defects, vacancies, and dislocations) among which, short-range scatterers play a more significant role in scattering at high carrier densities and are the main scattering source, limiting the graphene mobility at high carrier densities \cite{Sarma2011,Hwang2007}. 

We show that as carrier density increases, the interplay of the allowed scattering channels, enhanced screening, and chiral nature of the electronic structure determines the transport properties of the multiband scattering in multilayer graphene. Because the scattering rate is directly proportional to the density of states (DOS), and the DOS is enhanced at the bottom of the subbands, we find that the conductivity of multilayer graphene shows a sudden change when charge carriers begin to occupy the higher subbands. The change in conductivity arises mainly from the intersubband scattering due to the enhanced DOS at the band touching point. In particular, rhombohedral (periodic ABC) graphene shows a large conductivity drop when charge carriers fill the higher subband because of the diverging DOS at the bottom of the subbands. 

Using this unusual conductivity drop at the band touching point, we propose a novel design utilizing the large negative differential transconductance in multilayer graphene, in which the mean-free paths of charge carriers are controlled by gating \cite{Vaziri2013, Tse2008, Sakamoto2000, Palevski1989}. In addition, the conductivity drop in multilayer graphene could be used for electronic device applications such as amplifiers, oscillators, and multivalued logic systems. We also discuss the possibility of tuning the conductivity drop by changing the interlayer separation or doping method.

\begin{figure*}[t]
\includegraphics[width=0.9\linewidth]{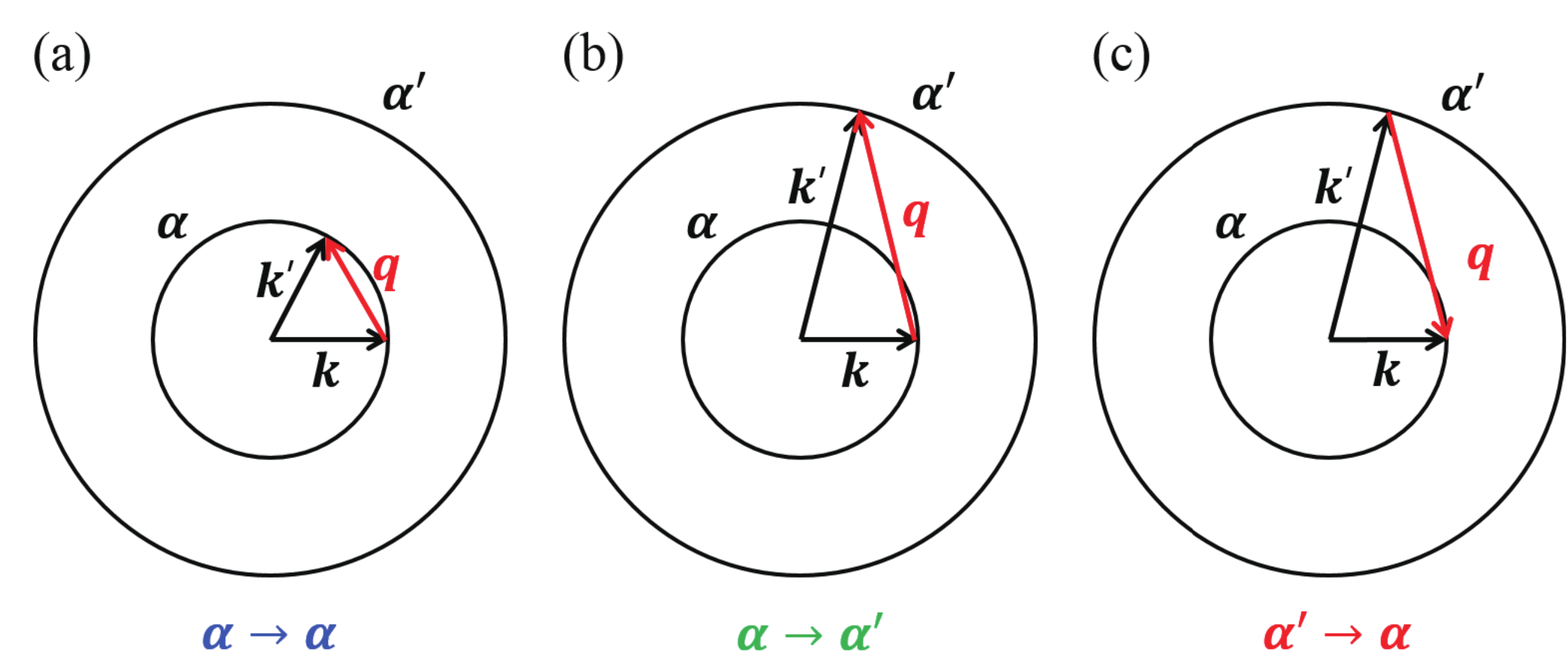}
\caption{Schematic picture of transition rates for (a) $P_{\alpha \rightarrow \alpha}^{(0)}$, (b) $P_{\alpha \rightarrow \alpha'}^{(1)}$ and (c) $P_{\alpha \leftarrow \alpha'}^{(2)}$. 
}  
\label{fig1}
\end{figure*}

\section*{METHODS}

In this paper, we calculate the density-dependent conductivity of multilayer graphene within the Boltzmann transport theory, focusing on the high density regime at which higher subbands are occupied by charge carriers and multiple subbands are involved in the scattering process. In our calculation, we incorporate the effects of multiband electronic scattering off the impurity centers, which are described by a set of coupled equations that relate the relaxation times for the multiple subbands involved in scattering \cite{Siggia1970,Sarma1987,Breitkreiz2013}.
Note that Boltzmann transport theory is known to be valid at the high density limit, and therefore, the multiband scattering can be described well in this semi-classical formulation.

Considering scattering processes involving multiple bands, the Boltzmann transport equation is given by \cite{Ashcroft}
\begin{equation}
\label{eq:Boltzmann}
(-e){\bm E}\cdot {\bm v}_{\alpha{\bm k}} \left(-\frac{\partial f_{\alpha{\bm k}}}{\partial\varepsilon_{\alpha{\bm k}}}\right) \!=\!\sum_{\alpha'} \!\int \!\! \frac{d {\bm k'}}{(2\pi)^2} W^{\alpha'{\bm k'}}_{\alpha{\bm k}} \,(f_{\alpha{\bm k}}-f_{\alpha'{\bm k'}})
\end{equation}
where $\bm{E}$ is an applied electric field, ${\bm v}_{\alpha{\bm k}}$ is the carrier velocity in the $\alpha$-th band with 2D momentum ${\bm k}$, $f_{\alpha{\bm k}}$ is the distribution function of the carriers in the $\alpha$-th band, $W^{\alpha'{\bm k'}}_{\alpha{\bm k}}={2\pi\over \hbar}\left|\left<\alpha \bm {k}\left|V_{\rm imp}(\bm{k}-\bm{k}')\right|\alpha' \bm{k'}\right>\right|^2 \delta(\varepsilon_{\alpha',\bm{k}'}-\varepsilon_{\alpha, \bm{k}})$ is the transition rate from $\alpha,{\bm k}$ to $\alpha',{\bm k'}$, and $V_{\rm imp}$ is the impurity potential. Note that $W^{\alpha'{\bm k'}}_{\alpha{\bm k}}=W^{\alpha{\bm k}}_{\alpha'{\bm k'}}$ because of the detailed balance. 
To solve the coupled Boltzmann transport equations, we can expand the distribution function up to the linear order of the electric field, i.e.,
$f_{\alpha\bm{k}}=f^{(0)}(\varepsilon_{\alpha\bm{k}}) + \delta f_{\alpha\bm{k}}$, where 
$f^{(0)}(\varepsilon)=[e^{(\varepsilon-\varepsilon_{\rm F})/k_{\rm B}T}+1]^{-1}$ is the equilibrium Fermi distribution function and $\delta f_{\alpha\bm{k}}$ is the non-equilibrium contribution proportional to the field.
By introducing energy-dependent transport relaxation time $\tau_{\alpha}$ associated with each band $\alpha$ and assuming $\delta f_{\alpha\bm{k}}=(-e){\bm E}\cdot {\bm v}_{\alpha{\bm k}} S^{(0)}(\varepsilon)\tau_{\alpha}$ at energy $\varepsilon=\varepsilon_{\alpha\bm{k}}=\varepsilon_{\alpha'\bm{k}'}$ where $S^{(0)}(\varepsilon)=-\frac{\partial f^{(0)}(\varepsilon) }{\partial\varepsilon}$, then Eq.~(\ref{eq:Boltzmann}) becomes
\begin{eqnarray}
\label{eq:Boltzmann2}
&&(-e){\bm E}\cdot \bm{v}_{\alpha \bm{k}} S^{(0)}(\varepsilon)\\
&=& \sum_{\alpha'}\int \!\! \frac{d^2 k'}{(2\pi)^2} W^{\alpha'{\bm k'}}_{\alpha{\bm k}}(-e){\bm E} 
\cdot\left[\bm{v}_{\alpha \bm{k}} \tau_{\alpha}-\bm{v}_{\alpha' \bm{k'}}\tau_{\alpha'}\right]S^{(0)}(\varepsilon). \nonumber 
\end{eqnarray}
Note that because of elastic scattering at energy $\varepsilon$, $S^{(0)}(\varepsilon)$ is cancelled in Eq.~(\ref{eq:Boltzmann2}). 
After matching coefficients in $\bm{E}$, we obtain the coupled equations for the relaxation time:
\begin{eqnarray}
\label{eq:multiband_scattering}
1=\tau_\alpha \left[P_{\alpha \rightarrow \alpha}^{(0)}+\sum_{\alpha'\neq\alpha}P_{\alpha \rightarrow \alpha'}^{(1)}\right]-\sum_{\alpha'\neq\alpha} P_{\alpha \leftarrow 
\alpha'}^{(2)}\tau_{\alpha'}.
\end{eqnarray}
The functions $P^{(i)}_{\alpha \rightarrow \alpha'}$ are the transition rates between bands $\alpha$ and $\alpha'$ (see Fig.~\ref{fig1}) given by
\begin{widetext}
\begin{subequations}
\begin{eqnarray}
\label{eq:intraband_transition}
P_{\alpha \rightarrow \alpha}^{(0)}(\bm{k})&=&\frac{2\pi}{\hbar}n_{\rm imp}\int \frac{d^2 k'}{(2\pi)^2}\left|V_{\rm imp}(\bm{k}-\bm{k}')\right|^2 F_{\alpha \bm{k}', \alpha \bm{k}}(\phi)  (1-\cos \phi)\, \delta\left(\varepsilon_{\alpha \bm{k}}-\varepsilon_{\alpha \bm{k}'}\right), \\
\label{eq:interband_transition1}
P_{\alpha \rightarrow \alpha'}^{(1)}(\bm{k})&=&\frac{2\pi}{\hbar}n_{\rm imp}\int \frac{d^2 k'}{(2\pi)^2}\left|V_{\rm imp}(\bm{k}-\bm{k}')\right|^2 F_{\alpha' \bm{k}', \alpha \bm{k}}(\phi)\, \delta\left(\varepsilon_{\alpha \bm{k}}-\varepsilon_{\alpha'\bm{k}'}\right), \\
\label{eq:interband_transition2}
P_{\alpha \leftarrow  \alpha'}^{(2)} (\bm{k})&=&\frac{2\pi}{\hbar}n_{\rm imp}\int \frac{d^2 k'}{(2\pi)^2}\left|V_{\rm imp}(\bm{k}-\bm{k}')\right|^2 F_{\alpha' \bm{k}', \alpha \bm{k}}(\phi) \frac{v_{\alpha'\bm{k}'}}{v_{\alpha\bm{k}}}\cos \phi\, \delta\left(\varepsilon_{\alpha \bm{k}}- \varepsilon_{\alpha' \bm{k}'}\right),
\end{eqnarray}
\end{subequations}
\end{widetext}
where $n_{\rm imp}$ is a randomly distributed impurity density, $\phi$ is an angle between the incoming and outgoing wavevectors of $\bm k$ and ${\bm k}'$, and $F_{\alpha' {\bm k}', \alpha {\bm k}}(\phi)=|\left<\alpha'{\bm k}'|\alpha{\bm k}\right>|^2$ is the square of the wavefunction overlap. Here the cosine weight factors represent the contribution from $\bm{v}_{\alpha' \bm{k}'}$ parallel to $\bm{v}_{\alpha \bm{k}}$ while that perpendicular to $\bm{v}_{\alpha \bm{k}}$ is cancelled in the $\bm{k}'$ integration. Thus, by solving the coupled equations in Eq.~(\ref{eq:multiband_scattering}) for multiband system, we can obtain the transport relaxation time $\tau_{\alpha}$ for each band \cite{Siggia1970,Sarma1987,Breitkreiz2013}.

If we consider a single-band system, only the intraband transition term $P_{\alpha \rightarrow \alpha}^{(0)}$ appears in Eq.~(\ref{eq:multiband_scattering}), and $P_{\alpha \rightarrow \alpha}^{(0)}$ becomes  $\tau_{\alpha}^{-1}$, i.e., the inverse of the one band momentum relaxation time, as in the well-known conventional relaxation time approximation \cite{Ashcroft}. In contrast, in a multiband system, $P_{\alpha \rightarrow \alpha'}^{(1)}$ represents the interband transition rate, which describes the scattering from bands $\alpha$ to $\alpha'$, whereas $P_{\alpha \leftarrow \alpha'}^{(2)}$ describes the interband transition rate from the $\alpha'$-th band scattered into the $\alpha$-th band, as schematically shown in Fig.~\ref{fig1}. 
Note that $P^{(i)}$ have different cosine weight factors, which leads to a reduced or enhanced contribution depending on the scattering directions. The weight factor $v_{\alpha' \bm{k}'}/v_{\alpha \bm{k}}$ in Eq.~(\ref{eq:interband_transition2}) arises from the velocity difference between bands $\alpha$ and $\alpha'$.

\begin{figure*}[t]
\includegraphics[width=1\linewidth]{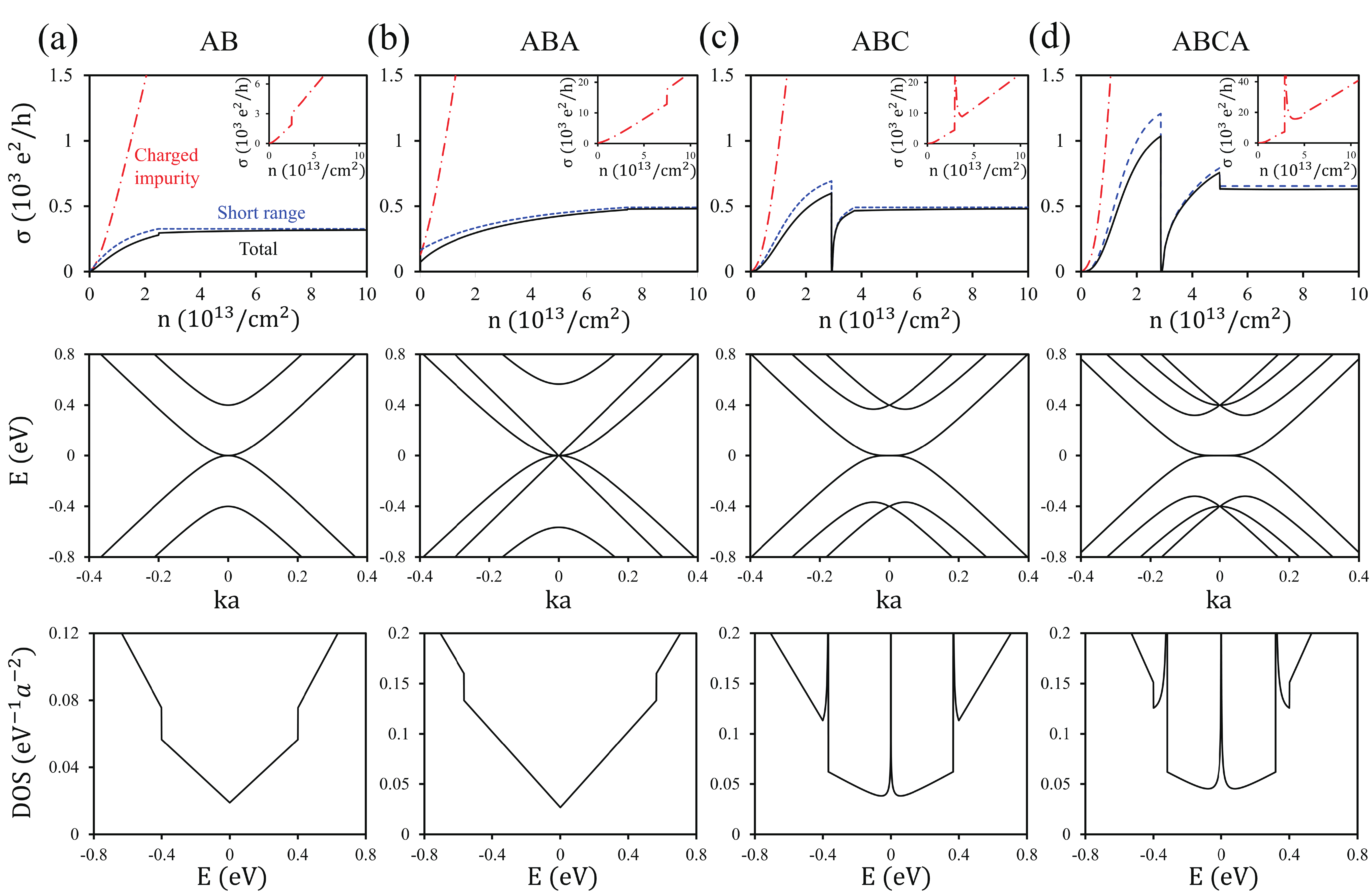}
\caption{Density dependence of conductivity (top panel), electronic structure (middle panel) and DOS (bottom panel) for (a) AB bilayer, (b) ABA trilayer, (c) ABC trilayer, and (d) ABCA tetralayer graphene. In the top panel, blue dotted and red dashed dotted lines represent the contributions from short-range scatterers and charged impurities, respectively, and black solid line represents the total conductivity. The insets show only the contribution from charged impurities.}
\label{fig:conductivity}
\end{figure*}

The current density induced by an electric field $\bm{E}$ is 
\begin{eqnarray}
\label{eq_current}
J_{i}=(-e)g_{s}g_{v}\sum_{\alpha}\int \frac{d^2 k}{(2\pi)^2} f_{\alpha\bm{k}} v_{\alpha\bm{k}, i} = \sum_{j}\sigma_{ij}E_{j},
\end{eqnarray}
where $g_{\rm s}$ and $g_{\rm v}$ are spin and valley degeneracies, respectively, and $\sigma_{ij}$ is the conductivity tensor given by 
\begin{eqnarray}
\label{eq_conductivity_tensor}
\sigma_{ij}=g_{\rm s} g_{\rm v} e^2  \sum_{\alpha}\int \frac{d^2 k}{(2\pi)^2} S^{(0)}(\varepsilon_{\alpha \bm{k}}) v_{\alpha\bm{k}, i}v_{\alpha\bm{k}, j}\tau_{\alpha \bm{k}}.
\end{eqnarray}
Assuming an isotropic band structure with $\sigma_{ij}=\sigma \delta_{ij}$, at zero temperature we finally obtain
\begin{eqnarray}
\label{eq_sigma}
\sigma=g_{\rm s} g_{\rm v} e^2 \sum_{\alpha} \rho_{\alpha}(\varepsilon_{\rm F}) {\cal D}_{\alpha},
\end{eqnarray}
where $\rho_{\alpha}(\varepsilon_{\rm F})$ is the DOS (per spin and valley) at the Fermi energy $\varepsilon_{\rm F}$, and ${\cal D}_\alpha=\frac{1}{2}v_{\alpha}^2\tau_{\alpha}$ is the diffusion constant of the $\alpha$-th band, where $v_{\alpha}$ and $\tau_{\alpha}$ are the Fermi velocity and transport relaxation time, respectively, at $\varepsilon_{\rm F}$. 

\begin{figure*}[t]
\includegraphics[width=1\linewidth]{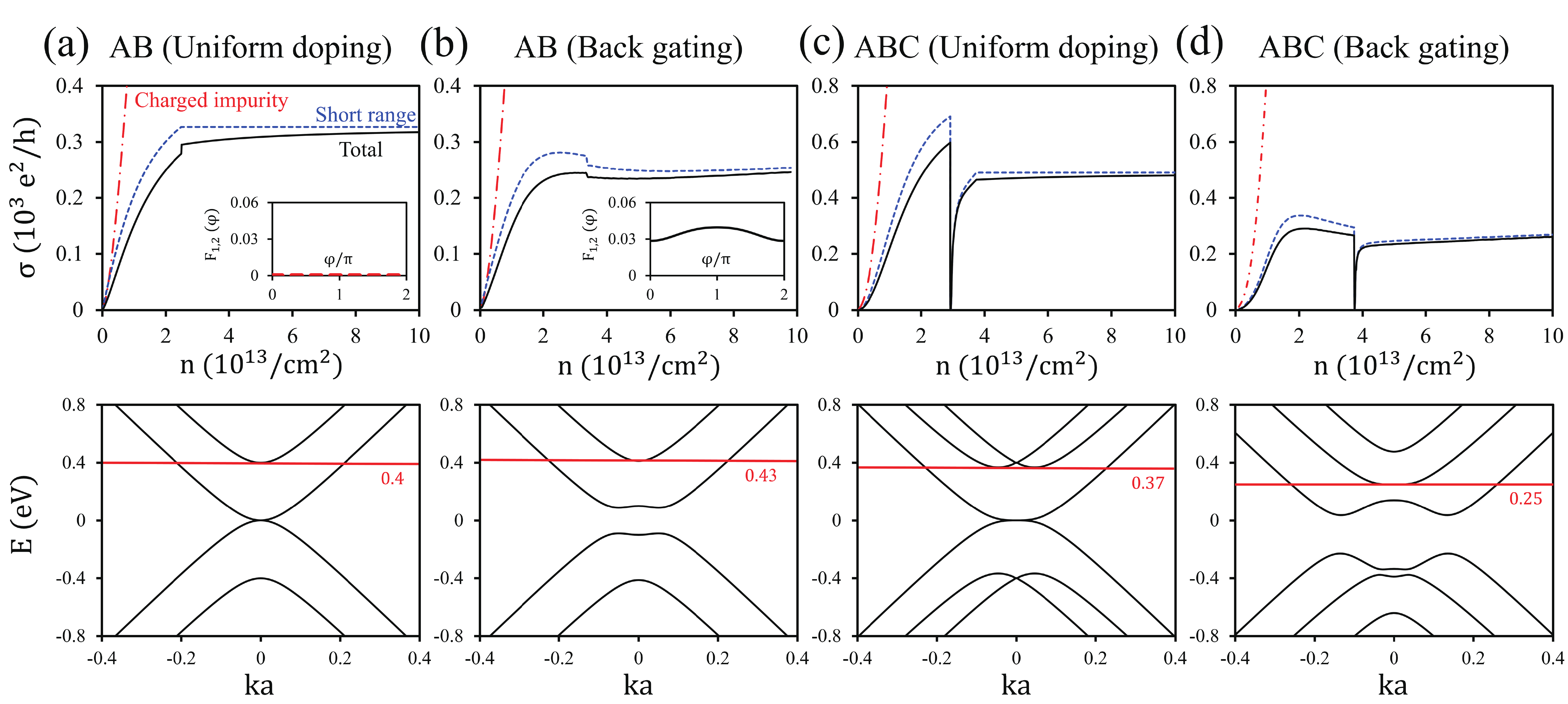}\caption{\label{fig:epsart}
Density dependence of conductivity (top panel) and electronic structure (bottom panel) for AB bilayer graphene with (a) uniform doping and (b) back gating, and for ABC trilayer graphene with (c) uniform doping and (d) back gating. Red lines in the bottom panel represent the band touching energy level for each doping. The insets in (a) and (b) show interband wavevector overlap $F_{1,2}(\phi)$ at the band touching point.}
\label{fig:external_field}
\end{figure*}

In this work, we consider two types of scattering sources: long-range Coulomb scattering and short-range scattering. In general, the Coulomb scattering is known to be an important scattering mechanism at low densities, whereas the short-ranged scattering mechanism is believed to be important at high carrier densities \cite{Sarma2011,Min2011}. The charged impurities are screened by the carriers, and we treat the screened Coulomb potential within the Thomas-Fermi approximation, $V_{\rm imp}^{\rm charge}({\bm q})=\frac{2\pi e^2}{\epsilon_0 (q+q_{\rm TF})}e^{-qd_{\rm imp}}$, where  $q_{\rm TF}=\frac{2\pi e^2}{\epsilon_0}D(\varepsilon_{\rm F})$ is the Thomas-Fermi wavevector, $\epsilon_0$ is the effective background dielectric constant, $D(\varepsilon_{\rm F})$ is the total DOS at the Fermi energy $\varepsilon_{\rm F}$ (which includes the spin and valley degeneracies as well as all the bands crossing $\varepsilon_{\rm F}$), and $d_{\rm imp}$ is the average distance between the impurities and the graphene sheet. The short-range scatterers describe atomic defects or vacancies, and can be approximated as a Dirac-delta function in the real space or a constant potential $V_{\rm imp}^{\rm short}({\bm q})=V_0$ in the momentum space. When we consider both charged impurities and short-range scatterers together, we add their scattering rates according to Matthiessen's rule, assuming that each scattering mechanism is independent.

From Eq.~(\ref{eq_sigma}), we first qualitatively describe the conductivity behaviors of multilayer graphene when higher subbands are occupied by charge carriers. When the Fermi energy reaches the higher subbands, there are three notable features in the transport properties. First, the conducting channels increase because charge carriers in the high-energy subbands participate in transport, which provides higher conductivity. Second, the scattering channels are increased because of the allowed interband scatterings, which make the scattering rate increase (or equivalently, conductivity decreases). Third, the screening is enhanced because the DOS increases because of the higher band contribution, which reduces scattering and increases conductivity for charged impurities. The net effect of the conductivity change is determined by competition among these effects \cite{Lange1993}.

\section*{RESULTS AND DISCUSSION}

In this section, we show the calculated conductivity of multilayer graphene in terms of the carrier density. 
In the calculations, we used the following parameters to qualitatively match with known experimental data \cite{Morozov2008,Ye2011}: $n_{\rm imp}^{\rm charge}=5\times10^{11}$ cm$^{-2}$, $\epsilon_0=2.25$, $d_{\rm imp}=0$ for charged impurities, and for short-range scatterers, $n_{\rm imp}^{\rm short} V_0^2=2.0$ (eV$\cdot\buildrel _\circ \over {\mathrm{A}}$)$^2$. For the tight-binding parameters, we used $t=3$ eV and $t_\perp=0.4$ eV for the nearest-neighbor intralayer and interlayer hopping terms, respectively, [see Fig.~\ref{fig0}(b)] neglecting remote hopping terms for simplicity \cite{Min2008a,Min2008b}.

Figure \ref{fig:conductivity} shows the numerically calculated conductivity as a function of carrier density for (a) AB bilayer, (b) ABA trilayer, (c) ABC trilayer, and (d) ABCA tetralayer graphene. A salient feature in our results is the conductivity drop at a certain carrier density ($n=2\sim 3\times 10^{13}$ cm$^{-2}$), where the second subband begins to be occupied by charge carriers. The conductivity drop is especially prominent in ABC and ABCA stacking.
In AB and ABA stacking, there is no significant change in conductivity at the bottom of the second subband, which can be attributed to the cancellation between the enhancement of interband scattering and the increase in conducting channels. We find that the conductivity drop is significant for the short-range scatterers, which are the dominant scattering source at high carrier densities. Thus, overall conductivity at high carrier densities follows the behavior of short-range scattering. When we consider only screened charged impurities, the conductivity shows a peak at the band touching point instead of a drop, which is mostly induced by the enhanced screening associated with the enhanced DOS. 

To understand these opposite behaviors at the band touching density for the two scattering sources, we consider the DOS dependence of the scattering rate for each impurity. Note that $P_{\rm\alpha \leftarrow \alpha'}^{(2)}$, which negatively contributes to the scattering rate, is typically much smaller than $P_{\rm\alpha \rightarrow \alpha'}^{(1)}$ because of the cosine weight factor in Eq. 4(c). The interband transition rate for the low-energy subband at the band touching point is then governed by the contribution from the low-energy subband to the high-energy subband, which is directly proportional to the higher band DOS, $P_{\rm low \rightarrow high}^{(1)} \sim \rho_{\rm high}(\varepsilon_{\rm F})\left|V_{\rm imp}\right|^2 $. Because $V_{\rm imp}$ is constant for short-range scatterers, the interband scattering rate becomes $P_{\rm low \rightarrow high}^{(1), {\rm short}}  \sim \rho_{\rm high}(\varepsilon_{\rm F})$, leading to the conductivity drop. Thus, the conductivity drop is significant in the ABC and ABCA stackings because of the divergent DOS at the band touching point. 
For screened charged impurities, the screening effects play a more important role in the scattering rate, and we have $P_{\rm \alpha \rightarrow \alpha'}^{(i), {\rm charge}} \sim {\rho_{\alpha'}(\varepsilon_{\rm F}) \over D^2(\varepsilon_{\rm F})}$ ($i=0,1,2$), where $D(\varepsilon_{\rm F})=g_{\rm s}g_{\rm v}\sum_{\alpha}\rho_{\alpha}(\varepsilon_{\rm F})$. Thus, the transition rate becomes approximately inversely proportional to the DOS, which decreases the overall transition rate. When the screening effect overwhelms the contribution from the interband transition in the scattering rate, the conductivity due to the charged impurities shows a peak at the band touching point. Because the scattering rates for short-range scatterers are much larger than those for charged impurities at the band touching point, the overall scattering rate of multilayer graphene is mainly governed by short-range scatterers and the total conductivity exhibits a drop at the band touching density. Note that at low carrier densities, the impurity scattering strongly depends on the stacking arrangements, whereas at high enough densities, multilayer graphene behaves as decoupled monolayer graphene sheets in which short-range scatterers dominate over charged impurities \cite{Min2011}. 

\begin{figure*}[t]
\includegraphics[width=1\linewidth]{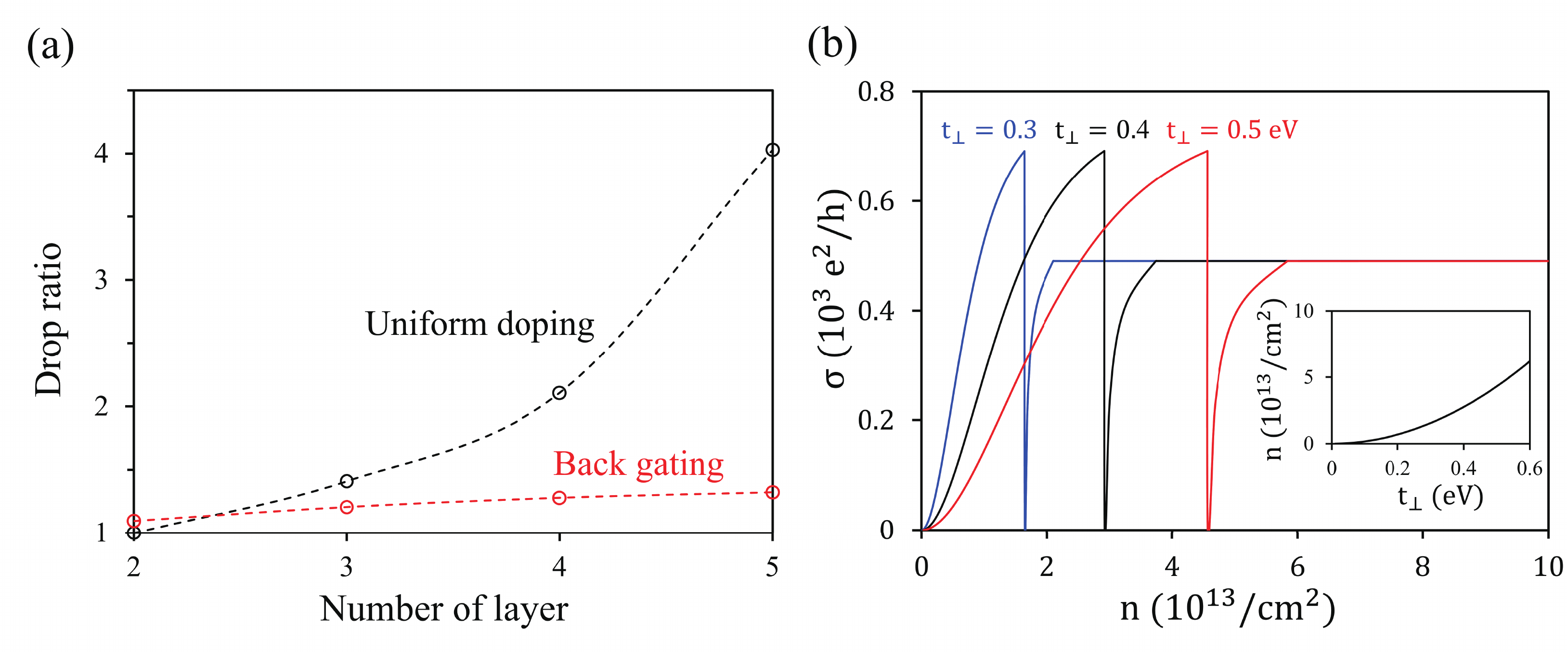}
\caption{(a) Conductivity drop ratio near the band touching point in rhombohedral stacking for uniform doping (black solid line) and back gating (red dashed line). (b) Density dependence of conductivity for several interlayer hopping terms $t_\perp$ in ABC stacking. Blue, black and red lines represent $t_\perp=0.3, 0.4, 0.5$ eV, respectively. The inset shows the band touching density as a function of $t_\perp$ for ABC stacking. 
}
\label{fig:interlayer_control}
\end{figure*}

The transport properties of the Bernal (periodic AB) and rhombohedral (periodic ABC) stacking arrangements are quite different because of the different electronic structures and chiral nature, even for the same number of layers, as shown in Fig.~\ref{fig:conductivity}. In general, rhombohedral stacking shows a more pronounced conductivity drop near the band touching density than Bernal stacking. Although Bernal stacking is more stable and common, rhombohedral stacking is also found in highly ordered pyrolytic graphite or natural crystal graphite at a concentration of approximately $15\%$ \cite{Lui2011}, and thus can be used for novel device applications, as is discussed later.  

In most graphene-based field-effect transistors, the carrier density is controlled by gate voltage and the electronic band structure is affected by the applied gate voltage. We now consider the electrically gated multilayer graphene in which charge carriers are supplied from the back gate and the layer charge is determined electrostatically under the boundary condition that the electric field above the top layer is zero.
Figures \ref{fig:external_field}(a) and (b) show the conductivity of AB bilayer graphene. We compare the result obtained without the band modification (uniform doping) with the result obtained with the gate-voltage dependent band change  (back gating). As shown in the insets of Figs.~\ref{fig:external_field}(a) and (b), the interband overlap factor $F_{1,2}(\phi)$ in back-gated AB bilayer graphene does not vanish, in contrast to the uniform doping case, which enhances the interband scattering, leading to a conductivity drop at the band touching point. In the case of ABC trilayer graphene, the minimum of the second subband in the presence of the gating occurs near ${\bm k}=0$ (but not exactly at ${\bm k}=0$ because of the formation of a Mexican hat structure in the trilayer and beyond), which reduces the DOS. Thus, the conductivity drop becomes smaller than that of the uniform doping case, as shown in Figs. \ref{fig:external_field}(c) and (d). 
These results indicate that the doping method affects the conductivity behavior at the bottom of the second subband by changing the energy band structure and its chiral nature. 

As discussed earlier, short-range scatterers are dominant over charged impurities at high carrier densities. In order to qualitatively understand the drop in conductivity, we focus on the density dependence of conductivity in the presence of dominant short-range scatterers only. 
We define the drop ratio as the ratio of conductivity just below the band touching density $n_{\rm touch}$ to that at $n=n_{\rm touch}+10^{13}$ cm$^{-2}$. In Fig.~\ref{fig:interlayer_control}(a), we show the conductivity drop ratio near the band touching point in rhombohedral stacking for uniform doping and back gating. The drop ratio increases with the number of layers because of the enhanced DOS at the band touching point. Note that the drop ratio for uniform doping is higher than that for back gating, which can be attributed to the occurrence of the band minima of the second subband away from ${\bm k}=0$, as shown in Figs.~\ref{fig:external_field}(c) and (d). 

Figure \ref{fig:interlayer_control}(b) shows the density dependence of conductivity for short-range scatterers for several values of the interlayer hopping $t_{\perp}$ in ABC graphene. 
It is possible to control the band touching density by adjusting the interlayer hopping constant $t_{\perp}$ through the interlayer separation between layers. 
Note that the interlayer hopping $t_{\perp}(d)$ with interlayer separation $d$ 
typically decays exponentially in the form of $t_{\perp}(d)=t_{\perp}\exp\left(-{d-d_0 \over r_0}\right)$, where $d_0=3.35$ $\mathrm \AA$ is the interlayer separation at which $t_{\perp}(d_0)=t_{\perp}$ and $r_0$ is the characteristic decay length for the hopping integral \cite{Slater1954, Koshino2013}.  
We find that the conductivity value just before the band touching ($\sigma_{\rm touch}\approx {\rm 25.38 } \frac{t^2 a^2} {n_{\rm imp}^{\rm short} V_0^2} \frac{e^2}{h}$,  where $a=2.46$ $\buildrel _\circ \over {\mathrm{A}}$ is the lattice constant) and the saturation value at large carrier densities ($\sigma_{\rm sat}= {\rm 18 } \frac{ t^2 a^2}{n_{\rm imp}^{\rm short} V_0^2}\frac{e^2}{h}$) do not depend on the interlayer hopping in the case of uniform doping. This means that a change in the interlayer hopping does not affect the drop ratio significantly, which is also true for back gating. As shown in the inset, however, the band touching density $n_{\rm touch}=\frac{25 t_{\perp}^2}{8\pi t^2 a^2}\propto t_{\perp}^2$ strongly depends on the interlayer hopping. Thus, the conductivity drop can occur at a much lower density if the interlayer separation is properly increased. We find that this trend is qualitatively true for a general rhombohedral graphene.

\section*{CONCLUSION}

In summary, we studied the electronic transport properties of multilayer graphene, focusing on the effect of multiband scattering at high carrier densities where the higher subband is occupied by carriers. 
By directly solving the coupled Boltzmann transport equations in the presence of both charged Coulomb impurities and short-range scatterers, we showed that the conductivity of multilayer graphene exhibits a sudden change when charge carriers begin to occupy the higher subbands, and thus a large negative differential transconductance (NDTC) appears as the carrier density varies. The large NDTC arises mostly from the intersubband scattering and the change in the DOS at the band touching point. We found that the conductivity drop in rhombohedral stacking is more prominent than that in Bernal stacking because of divergent DOS at the bottom of the higher subbands and increased interband scatterings. We also showed that the conductivity drop is affected by the doping method and interlayer separation, and an efficient conductivity drop can be obtained for uniformly doped rhombohedral graphene. 
Based on our results, it may be possible to design novel devices utilizing the large NDTC in multilayer graphene in which the mean-free paths of the charge carriers are controlled by gating. In addition, the conductivity drop in multilayer graphene could be used for electronic device applications such as amplifiers, oscillators, and multivalued logic systems. 
For  better device applications, we propose decreasing the band touching density by changing the interlayer separation and increasing the conductivity drop using uniform doping in rhombohedral stacking.

\section*{ACKNOWLEDGMENTS}

This research was supported by Basic Science Research Program through the National Research Foundation of Korea (NRF) funded by the Ministry of Education under Grant No. 2015R1D1A1A01058071 (S.W. and H.M.) and 2017R1A2A2A05001403 (E.H.H.).

\end{document}